\documentclass[twocolumn,showpacs,amsmath,amssymb]{revtex4}
\usepackage{graphicx}
\usepackage{bm}
\begin{document}
\title{Ferromagnetism from localized deep impurities in magnetic semiconductors.}
\author{Victor Barzykin} 
\affiliation{Department of Physics and Astronomy, University of Tennessee,
Knoxville, TN  37996-1200}
\begin{abstract}
We propose that localized defects in magnetic semiconductors  
act as deep impurities and can be described by the 
Anderson model. Within this model, hybridization of 
d-orbitals and p-orbitals gives rise to a non-RKKY indirect exchange mechanism, 
when the localized d-electrons are exchanged through both conduction 
and valence bands. For semiconductors with indirect band gap the non-RKKY part of exchange integral is
antiferromagnetic, which suppresses ferromagnetism.
In case of direct band gap, this exchange mechanism can, under certain conditions, lead to 
enhancement of ferromagnetism.
The indirect exchange intergral is much stronger than RKKY, and can be sufficiently long range. 
Thus, a potentially new class of high-temperature magnetic semiconductors emerges, where
doped carriers are not necessary to mediate ferromagnetism. 
Curie temperatures in such magnetic semiconductors are determined mostly by the interaction between 
localized impurities, \textit{not Zener mechanism}. This effect could also
be responsible for unusually high Curie temperatures in some magnetic semiconductors with
direct band gap, such as Ga$_{1-x}$Mn$_x$As.
\end{abstract}
\vspace{0.15cm}

\maketitle
 
\section{Introduction}

The advantage of ferromagnetic semiconductors (FS) as a source of 
spin-polarized carriers is that they can be easily integrated into semiconductor devices\cite{Prinz, Wolf}.
When discovered, ferromagnetism at room temperatures with full polarization of itinerant carriers will be
a major breakthrough in semiconductor electronics. Most theoretical
and experimental efforts have been concentrated on III-V, group IV, and II-VI-based diluted magnetic semiconductors (DMS).
These semiconductors are alloys in which some atoms are randomly replaced by magnetic atoms, such as Mn$^{2+}$.

Ferromagnetism in diluted magnetic semiconductors (DMS) is thought to be well understood in terms of the 
so-called p-d exchange model,
which was first considered  
over 50 years ago\cite{Zener1,Zener2,AG} (See also Ref.\cite{Korenblit, Furdyna, Dietll} for a review). 
At concentrations of impurities above the Mott limit, i.e., as soon as carriers become delocalized,
conventional model of FS is fairly simple.
According to, for example, Ref.\cite{AG}, the interaction between
charge carriers in a semiconductor and spin-S impurities can be written as:
\begin{eqnarray}
\label{int}
U &=& - \int \bm{s}(\bm{r}) \sum_i \bm{S}_i J^{pd}(\bm{r} - \bm{R}_i) d^3 \bm{r} \\ 
  &=& - J^{pd}_{\bm{q}=0} \bm{s} \sum_i \bm{S}_i, \nonumber
\end{eqnarray}
where $\bm{S}_i$ and $\bm{R}_i$ are the spin and the position of an i-th atom of magnetic impurity, 
$J^{pd}_{\bm{q}=0} = \int J(\bm{r}) d^3 \bm{r}$, and we have used the fact that magnetic impurities are randomly distributed
in the sample. A simple analysis\cite{AG} then shows that, in the presence of one type of charge carriers, Curie temperature,
$T_c$, is proportional to concentration of magnetic impurities, $N_i$, and the square of the strength of the exchange 
interaction, $a$:
\begin{equation}
T_c = {n_i S (S+1) (J^{pd}_{\bm{q}=0})^2 \chi_0 \over 12 \mu_0^2}
\label{Curie}
\end{equation}
Here $\mu_0$ is the magnetic moment of charge carriers, $\chi_0$ is the Pauli term in the spin susceptibility
in the absence of impurities. At sufficiently large carrier densities, 
the spin polarization of the charge carriers at $T=0$ is given by:
\begin{equation} 
s = {n_i J^{pd}_{\bm{q}=0} S \chi_0 \over 4 \mu_0^2}
\end{equation} 
In general, for any carrier density, the spin polarization is given by Ze\`eman-split Fermi surface,
with the difference in chemical potentials for ``up'' and ``down'' spins given by:
\begin{equation}
\mu_{\uparrow} - \mu_{\downarrow} = J^{pd}_{\bm{q}=0} n_i S
\end{equation}
The spin polarization at any filling is then very easily calculated from this equation. For example
when the carrier density 
\begin{equation}
n_e \le n_c = \frac{(2 J^{pd}_{\bm{q}=0} n_i S m^*)^{3/2}}{6 \pi^2}\, ,
\end{equation}
the carriers will be fully polarized at $T=0$. When $n_e > n_c$, they are no longer fully polarized.
The polarization of carriers is then determined by a parametric equation 
(with $\mu$ as a parameter):  
\begin{equation}
s = \frac{1}{2}\, (n_{e \uparrow} - n_{e \downarrow}),
\end{equation}
where   
\begin{equation}
n_{e \uparrow, \downarrow} = 
\frac{(2 m^* \mu \pm  J^{pd}_{\bm{q}=0} n_i S m^*)^{3/2}}{6 \pi^2}\,,
\end{equation}
and $n_e = n_{e \downarrow} + n_{e \uparrow}$.

Despite the simplicity of the basic concept of Zener ferromagnetism, calculations of $T_c$ for real materials 
become rather involved, and depend crucially on details of the
band structure, the $p-d$ exchange matrix, or direct antiferromagnetic exchange between $Mn^{2+}$ spins.
Most theoretical and experimental efforts on magnetic semiconductors have been concentrated on  
finding a new DMS-based material that would have a ferromagnetic transition above room temperature, and would be possible to
incorporate in thin film form with mainstream semiconductor device materials.    
There are theoretical predictions for T$_c$'s above room temperatures in several classes of these materials\cite{dietl1,dietl2}.  
However, experiments indicate that the growth of Curie temperature with concentration of magnetic impurities
saturates at 5-10\% Mn doping in most ferromagnetic  semiconductors; it may even start to decrease at higher
concentrations of Mn. Since, according to the mean field p-d model, the Curie temperature Eq.(\ref{Curie}) 
grows linearly or faster (if the carrier concentration changes) with the growth of Mn concentration, it is
important to understand what limits the growth of $T_c$, and how this can be avoided. 

Experimental  effort in FS concentrated mostly on In$_{1-x}$Mn$_x$As\cite{Munekata} ($T_c \sim 35K$) 
and Ga$_{1-x}$Mn$_x$As\cite{deBoeck,Ohno,Ohno1,Ohno2} ($T_c \sim 110K$). 
While room temperature ferromagnetism in FS remains a theoretical possibility, the highest Curie temperature reported in
the homogeneous sample with Mn concentration around 5\% is 116K for Mn$_x$Ge$_{1-x}$ \cite{Park}. 
The saturation $T_c$ growth with increased Mn concentration is usually 
ascribed to increased disorder. Disorder effectively introduces an exponential cutoff for the RKKY interaction and 
reduces its range\cite{Ohno}. However, as we discuss in section VIII, weak disorder does not change Curie temperature. 
Another possible reason is that
the strength of direct antiferromagnetic exchange grows as the average distance between impurities becomes shorter,
which, in turn, lowers the Curie temperature. 
Room temperature ferromagnetism has been observed in a number of compounds with \textit{large} concentration of $Mn$
impurities (see Ref. \cite{Wolf} for a review), or, due to phase separation and formation of nanoclusters of these compounds (
such as Mn$_{11}$Ge$_8$ in Ref.\cite{Park}).
However, large carrier concentration in these compounds limits the degree of spin polarization that is necessary
for device applications.

   Some theoretical studies\cite{Millis} claimed that the reason Curie temperature can't get higher above 
certain concentration of Mn is that the $p-d$ model
cannot be treated in the mean field approxination. However, since the RKKY interaction has a large radius, 
mean field treatment in absence of strong disorder should be justified\cite{Larkin}.

In what follows, we adopt the idea that the reason for the discrepancy between mean field theory and experiment lies
in the presence of additional interactions in the effective spin Hamiltonian of the problem, which are not present
in the conventional formulation of the $p-d$ model. To derive effective spin Hamiltonian, including additional interactions, 
we start from a more general Anderson model of deep magnetic impurities in FS.
We show that, in case of a deep Anderson impurity in a semiconductor,
an additional long-range indirect exchange interaction appears in the effective Hamiltonian, which, if 
antiferromagnetic,  severely limits $T_c$-s in these materials. On the other hand, for direct gap semiconductors, this  
interaction could change sign and become ferromagnetic, if the magnetic impurity lies deep enough. 
We derive conditions under which this becomes possible. As a result, ferromagnetic correlations would become
enhanced, not reduced, as the concentration of magnetic impurities grows, and high-temperature ferromagnetism could be 
possible even without carriers. 
The indirect exchange between two deep impurities, whether ferro- or antiferromagnetic, is stronger than the RKKY interaction, 
and thus could produce large Curie temperatures. For example, it could provide a possible explanation of high-temperature 
weak ferromagnetism\cite{Young} in La$_x$Ca$_{1-x}$B$_6$ or recently discovered 
CaB$_2$C$_2$\cite{Akimitsu} (CaB$_2$C$_2$ has $T_c=770K$ and $M=10^{-4} \mu_B$).
Experiment\cite{Akimitsu,Fisk} indicates that these materails are direct band semiconductors with a relatively
small band gap, and that impurities play a major role in establishing the new high-temperature
ferromagnetic state. However, other reports\cite{Sato} claim that high-temperature ferromagnetism in these 
materials is not a bulk effect. Rather, it is related to clustering
or new boron phases with Fe or Ni magnetic impurities. Understanding
the mechanism of weak ferromagnetism in these materials could lead to a discovery of more members of this class. 
In either case, as we report in this paper, there is a theoretical
possibility of a new type of FS with direct band gap, where ferromagnetism is not carrier-driven. 

This paper is organized as follows. In Section II we discuss the general procedure of the derivation of the effective
low-energy Hamiltonian for Anderson impurities. In section III we discover that, due to strong hybridization,
an Anderson impurity is no longer a local center; its localized wave function acquires a finite range, 
which is directly related to the corresponding "Bohr" radius of a charged impurity. Sections IV and V are devoted to 
the derivation of the effective exchange Hamiltonian for magnetic impurities. In Section VI we
explore the consequences of the large range of interaction for magnetism, such as a high Curie temperature in case of a direct band
gap if magnetic impurities are dense enough. In section VII we consider a minor modification of the effective Hamiltonian in 
case of higher spin (such as $S=5/2$ for Mn), and the application of these ideas to Ga$_{1-x}$Mn$_x$As. 
In section VIII the influence of weak disorder and interactions on Curie temperature is briefly discussed.
Section XI provides a summary and conclusions.

\section{The effective Hamiltonian}

   We start by considering a simple model of FS. In III-V systems, such as Ga$_{1-x}$Mn$_x$As, 
it is well established that the Mn ions substitute for Ga, and contribute itinerant holes to the GaAs  valence band. Experimentally,
the hole density is typically a small fraction (15\% or so) of the Mn concentration, perhaps due to strong localization of carriers
on Mn and other defects, so Ga$_{1-x}$Mn$_x$As can be considered partially compensated. The Mn ion has half-filled d-shell, which
acts like a spin-5/2 local moment. The Anderson model, which is more general 
than the $p-d$ model, should completely account for all the physics of FS. It is well known that, when spins are well localized,
the single-impurity Anderson Hamiltonian is reduced to the $p-d$ Hamiltonian by the Schrieffer-Wolff transformation\cite{SW}.
For many impurities, this may no longer be the case. Let us start by considering a single-orbital Anderson Hamiltonian:

\begin{equation}
H = H_0 + H_V,
\label{theH}
\end{equation}
where
\begin{eqnarray}
H_0 &=& \sum_{\bm{p} \sigma i} \epsilon_i(p) a^{\dagger}_{i \bm{p} \sigma} a_{i \bm{p} \sigma} + \\
&+& \sum_n \left[ \epsilon_0 \sum_{\sigma}  d^{\dagger}_{n \sigma} d_{n \sigma} + 
U d^{\dagger}_{n \uparrow} d^{\dagger}_{n \downarrow} d_{n \downarrow} d_{n \uparrow} \right]. \nonumber
\end{eqnarray}
   
Here the first sum ($\bm{p}$) is taken over the reciprocal space, the second sum ($n$) over real space impurity sites.
$U$ is the on-site Coulomb repulsion term. Typically, $U$ is very large ($\sim 5 eV$), and can be taken to be infinite.
Here $\epsilon_i(p)$ are the energy band spectra of conduction and valence bands ($i=1,2$).

The hybridization term in the model Hamiltonian, $H_V$, accounts for the $p-d$ hybridization between impurity sites and 
conduction and valence bands:

\begin{equation}
H_V = \frac{1}{N^{1/2}} \sum_{\bm{p} n \sigma i} V_i \left\{ a^{\dagger}_{\bm{p} \sigma i} d_{n \sigma} e^{-i \bm{p} \cdot \bm{R}_n}
+ h.c.\right\}
\end{equation}

This model is a reasonable generalization of the $p-d$ exchange Hamiltonian, usually considered in the literature. 
Because of large on-site Coulomb
repulsion, the $d$-levels are half-filled. While we consider the case of a single d-orbital, a generalization 
to $S=5/2$ Mn ion is straightforward (see section VII below). 

The Anderson Hamiltonian Eq.(\ref{theH}) describes a very complicated problem.
However, under the $U=\infty$ constraint, it can be reduced to the problem of Heisenberg spins 
(in case of a single d-orbital, spin-$1/2$).
The low-energy effective 
Hamiltonian is equivalent to Eq.(\ref{theH}) in the limit $k_B T \ll \Delta_i$, where $\Delta_i$ is the energy difference
between the impurity $d$-level and the top of the valence band, or the energy difference between the impurity $d$-level and the 
bottom  of conduction band. Since typical gap values in semiconductors are of the order of $1 eV$, this is usually a valid assumption.

\begin{figure}
\includegraphics[width=3in]{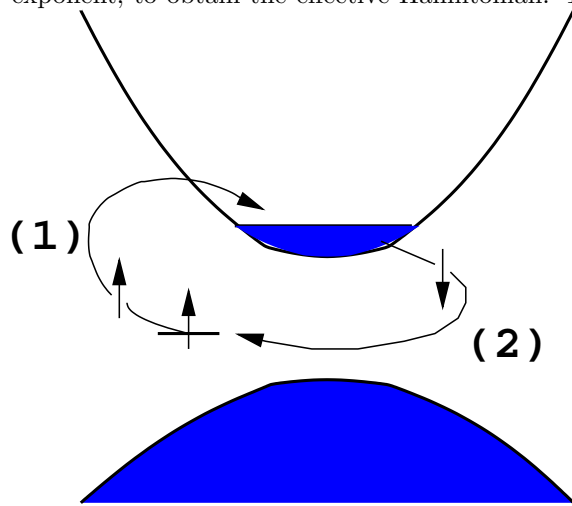}
\caption{The second-order contribution to the effective Hamiltonian, shown as a time-ordered process.}
\label{fig2ord}
\end{figure}

Following Refs. \cite{GS1,BG1}, the effective spin Hamiltonian can be derived by expanding the S-matrix
(or, at finite temperatures, the partition function) in $V$-s, and re-expressing various time-ordered processes
in terms of spin operators. Then, these processes are collected back under exponent, to obtain the effective
Hamiltonian. This method allows one to obtain 
consistently the interaction between spins and carriers (electrons or holes) in conduction and valence bands, and 
carrier-carrier interaction. In a way, the concept is similar to perturbative renormalization group, since 
we arrive at a low energy effective Hamiltonian by integrating out higher-energy states. Treating hybridization term in Eq.(\ref{theH}) 
as a perturbation, we can rewrite the partition function as 

\begin{equation}
Z = Tr[\exp(-\beta H_0) S(\beta)],
\end{equation}
where
\begin{widetext}
\begin{equation}
S(\beta) = T \exp\left( - \int_0^{\beta} H_V(\tau) d \tau \right) =
T \exp(- \sum_{n \sigma i} \int_0^{\beta} d \tau V_i \left\{ \Psi^{\dagger}_{n \sigma i}(\tau) d_{n \sigma}(\tau) 
+ d^{\dagger}_{n \sigma}(\tau) \Psi_{n \sigma i}(\tau)\right\}
\label{smatr}
\end{equation}
\end{widetext}

The problem of finding the effective Hamiltonian is then to reduce these expressions to the form:

\begin{equation}
Z = Tr(\exp[-\beta H_{eff}]),
\end{equation}
using the $k_B T \ll \Delta_i$ condition, i.e., the fact that the local levels are almost always occupied, and that transitions to
conduction and valence bands are absent at low temperatures. Various terms in the effective Hamiltonian can then
be associated with certain time-ordered virtual processes.
For example, the first non-zero contribution to $H_{eff}$ is from the second order term in the expansion of 
Eq.(\ref{smatr}) in $V_i$:
\begin{widetext}
\begin{equation}
S_2(\beta) = \sum_{n \sigma \sigma'} V_1^2 \int_0^{\beta} d \tau_1
\int_{\tau_1}^{\beta} d \tau_2 \Psi_{n \sigma 1}(\tau_2) \Psi^{\dagger}_{n \sigma' 1}(\tau_1)
d^{\dagger}_{n \sigma}(\tau_2) d_{n \sigma'} (\tau_1).
\label{2order}
\end{equation}
\end{widetext}
The order of operators $d^{\dagger}_{n \sigma}(\tau_2) d_{n \sigma'} (\tau_1)$ ($\tau_2 > \tau_1$) is fixed
by the assumption of strong Coulomb repulsion on the $n$-th center. Note that, in the second order, because
all centers are filled, only one band contributes to the effective Hamiltonian. The "filled" band does not contribute,
because of the Pauli principle (this is not the case at a finite U). The perturbative time-ordered process corresponding 
to the term in Eq.(\ref{2order}) is shown in Fig.\ref{fig2ord}.

We may rewrite $\Psi_{n \sigma 1}(\tau_2) \Psi^{\dagger}_{n \sigma' 1}(\tau_1)$ in the interaction representation
as:
\begin{widetext}
\begin{equation}
\Psi_{n \sigma 1}(\tau_2) \Psi^{\dagger}_{n \sigma' 1}(\tau_1) = 
\{\Psi_{n \sigma 1}(\tau_2) \Psi^{\dagger}_{n \sigma' 1}(\tau_1)\}_+ - 
\Psi^{\dagger}_{n \sigma' 1}(\tau_1) \Psi_{n \sigma 1}(\tau_2). 
\label{commut}
\end{equation}
\end{widetext}
The first term in Eq.(\ref{commut}) is a c-number. In the limit $\Delta_1 \gg k_B T$ it is
possible to put $\tau_1 \simeq \tau_2$ in the second term in Eq.(\ref{commut}). Thus, for
the second-order contribution in the effective Hamiltonian we may write:
\begin{widetext}
\begin{equation}
H^{(2)}_{eff} = - \sum_{\bm{p}} \frac{V_1^2}{\epsilon_{1p} - \epsilon_0}\, 
+ \frac{V_1^2}{\Delta_1}\, \sum_{n \sigma \sigma'} \Psi^{\dagger}_{n \sigma' 1} \Psi_{n \sigma 1}
\left( \frac{1}{2}\, \delta_{\sigma \sigma'} + \bm{S}_n \bm{\sigma}_{\sigma' \sigma} \right),
\label{2oref}
\end{equation}
\end{widetext}
where $\bm{S}_n$ is the localized spin of the n-th impurity. The first term corresponds to the renormalization of 
the energy of the localized level, while the second term involving spins of localized impurity
and carriers is nothing but the ordinary $H_{pd}$, the p-d model discussed in section I.

\begin{figure}
\includegraphics[width=3in]{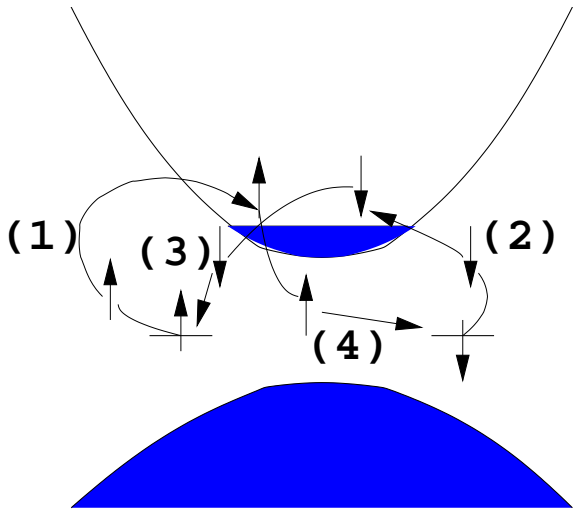}
\caption{The fourth-order antiferromagnetic superexchange contribution to the effective exchange interaction between two localized 
impurities.}
\label{fig4orda}
\end{figure}

\begin{figure}
\includegraphics[width=3in]{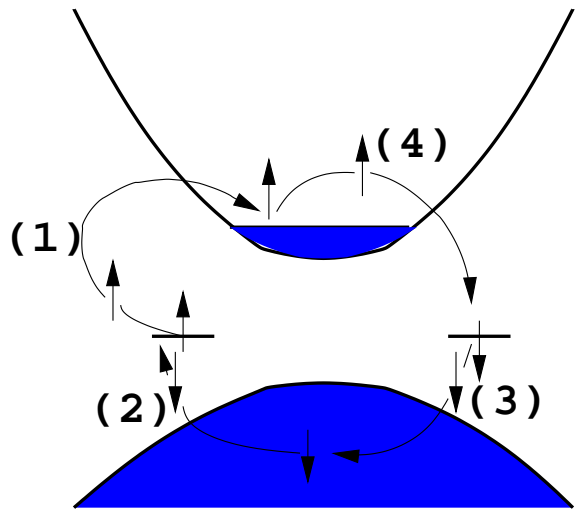}
\caption{The fourth-order Bloembergen-Rowland term in the effective exchange interaction.} 
\label{fig4ordb}
\end{figure}

The next order terms in the effective Hamiltonian are fourth order in V-s. There are sums
over two local centers, $m$ and $n$ in $S_4(\beta)$. Also, there are contributions to $S_4(\beta)$,
which we will denote $S'_4(\beta)$, that are already accounted for in the effective 
Hamiltonian Eq.(\ref{2oref}):
\begin{equation}
S'_4(\beta) = \frac{1}{2}\, \int \int d \tau_1 d \tau_2 T\{H_{pd}(\tau_1) H_{pd}(\tau_2)\}.
\label{subtr}
\end{equation} 
These contributions need to be subtracted from $S_4(\beta)$, to get the fourth-order (in $V_i$)
contributions in the effective Hamiltonian. The most important contribution 
is the effective exchange interaction between localized spins:
\begin{equation}
H^{(4)}_{ex} = - \sum_{n \neq m} J(\bm{R}_n - \bm{R}_m) \bm{S}_n \bm{S}_m
\label{exchangeint}
\end{equation}
This exchange interaction is the result of 2 time-ordered processes shown in Figs. \ref{fig4orda}, \ref{fig4ordb}.
One is the superexchange (Fig. \ref{fig4orda}), which is a result of the localized spins exchanged
through the empty conduction band. The other process is the Bloembergen-Rowland term\cite{BR} (Fig. \ref{fig4ordb}),
an exchange process through both conduction and valence bands. The form of these contributions will be 
discussed in greater detail in sections IV and V below.

In addition, other interesting contributions arise as a result of fourth-order processes, such as p-d scattering by
spins on two centers,
\begin{widetext}
\begin{equation}
H^{(4)}_{pd2c} = \frac{V_1^2}{\Delta_1^2} \sum_{n \neq m, \sigma \sigma'} t(\bm{R}_n - \bm{R}_m)
\Psi^{\dagger}_{m \sigma' 1} \Psi_{n \sigma 1} [\bm{S}_n \bm{S}_m \delta_{\sigma' \sigma} + 
i ([\bm{S}_m \bm{S}_n] \hat{\bm{\sigma}}_{\sigma' \sigma})], 
\end{equation}
\end{widetext}
nontrivial local contribution,
\begin{widetext}
\begin{equation}
H^{(4)}_{local} = - \frac{V_1^4}{\Delta_1^3} \sum_{n \sigma \rho \rho'}
\Psi^{\dagger}_{n \sigma 1} \Psi^{\dagger}_{n \rho' 1} \left[\frac{1}{2}\, \delta_{\rho' \rho}  +
(\hat{\bm{\sigma}}_{\rho' \rho} \bm{S}_n)\right] \Psi_{n \rho 1} \Psi_{n \sigma 1},
\end{equation}
\end{widetext}

and corrections to the energy of the local level and ground state energy, which are dropped.
These nontrivial terms, however, are higher order in carrier density, which is small in magnetic
semiconductors. Thus, they don't play any significant role in magnetism, and, hence, will not be discussed in
any detail. We refer the reader to references \cite{GS1, GS2, BG1}, where these terms were discussed in detail
in connection with the 3-band model of cuprate superconductors.

\section{Renormalization of local level due to hybridization.}

To clarify the new physics of an Anderson impurity, in comparison with 
$p-d$ magnetic impurity, let us first consider a single-level problem ($\bm{R}_n = 0$),
interacting only with the conduction band (i.e., we take $V_2=0$ for simplicity).
The Hamiltonian is defined on the manifold of wavefunctions $\{\Phi_0, \Psi_{1 \bm{p_1}},
\Psi_{1 \bm{p_2}}, ... \}$, which correspond to the
exact solution for a center in crystalline lattice in the absence of hybridization. These
wavefunctions form a complete orthogonal basis for the single-particle problem. We may now
introduce the hybridization as a perturbation into the Schr\"odinger equation. In the matrix representation,
the single-level Hamiltonian takes a very simple form:
\begin{equation}
H = \left\{ \begin{array}{cccc}
\epsilon_0 & v_1 & v_1 & \cdots \\
v_1 & \epsilon_{1 \bm{p}_1} & 0 & \cdots \\
v_1 & 0 & \epsilon_{1 \bm{p}_2} & \cdots \\
\vdots & \vdots & \vdots & \ddots 
\end{array} \right\}
\end{equation}
Here $v_1 \equiv V_1/N^{1/2}$. This Hamiltonian can be 
easily diagonalized, by looking for a solution for the ground state wave function as a linear
combination of all single-particle states:
\begin{equation}
\Psi = \chi_0 \Phi_0 + \sum_{\bm{p}} \{\chi^{(1)}_{\bm{p}} \Psi_{1 \bm{p}}\},
\end{equation}
where the coefficients obey the following set of equations:
\begin{eqnarray}
\label{spectrum}
(\epsilon_0 - \epsilon) \chi_0 + v_1 \sum_{\bm{p}} \chi_{\bm{p}} &=& 0 \\
 v_1 \chi_0 + (\epsilon_{1 \bm{p}} - \epsilon) \chi_{\bm{p}}&=& 0 \nonumber
\end{eqnarray}

\begin{figure}
\includegraphics[width=3in]{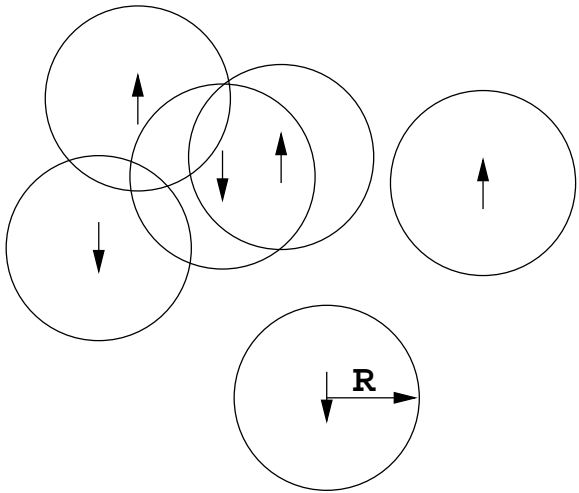}
\caption{Localized centers acquire a finite radius in the Anderson model. This radius can be as large
as 2-3 lattice spaces in magnetic semiconductors. An overlap of wavefunctions from two localized level
leads to the new physics, which is not captured by the traditional $p-d$ model.}
\label{fig3}
\end{figure}

Solving Eq.(\ref{spectrum}) gives the new position of the localized level:
\begin{equation}
\epsilon = \epsilon_0 - \frac{V_1^2}{N}\, \sum_{\bm{p}} \frac{1}{\epsilon_{1 \bm{p}} - \epsilon}
\label{level}
\end{equation}
Thus, in presence of hybridization the position of the impurity level is shifted. If the depth
of the level $\Delta_1 \ll D_1$, where $D_1$ is the bandwidth of the conduction band (which
is usually a valid assumption for semiconductors, where the band gap is much smaller than the
corresponding bandwidths of conduction and valence bands), the sum in Eq.(\ref{level}) can
be performed numerically. The correction to the energy level,
\begin{equation}
\epsilon - \epsilon_0 \sim \frac{V_1^2}{D_1},
\end{equation}
turns out to be small for small enough
hybridization parameter $V_1 \ll \sqrt{D_1 \Delta_1}$. It is responsible for a
slight downward shift of the energy level. Such corrections will thus be dropped
in our further discussion.

The wavefunction of the localized level, according to Eq.(\ref{spectrum}),
acquires an admixture to its decay in the form:
\begin{equation}
\delta \Phi_0(r) \sim V_1 m_1 a^2 \frac{a}{r}\, e^{-r \sqrt{2 m_1 \Delta_1}}.
\label{wavef}
\end{equation}
Here $a$ is the lattice constant (we assume a simple cubic lattice), $m_1$ is the effective mass 
of carriers in the empty band (for hole-doped materials, the valence band is the ``empty'' band, 
since it is empty of holes), $\Delta_1$ is the energy difference between impurity levels and the bottom 
of the empty band. 
Thus, a new length scale enters the problem,
\begin{equation}
R_0 = (2 m_1 \Delta_1)^{-1} \sim a \sqrt{D_1/\Delta_1},
\label{scale}
\end{equation}
which, while small compared to the average distance between doped carriers, may
considerably exceed interatomic distances (the large parameter is the square root of
the ratio of the bandwidth to the energy gap in a semiconductor), and become comparable to the average
distance between magnetic impurities. The overlap of the localized wave functions
gives rise to a new contribution to the exchange integral, which will be derived
below. The picture of finite-radius localized magnetic impurities with overlapping wavefunctions is
shown in Fig.\ref{fig3}.

Existence of a large (compared to lattice spacing) radius for deep impurities was first noticed by
Keldysh\cite{Keldysh}, who analyzed deep charged impurities in ordinary semiconductors.
In his case, however, this scale corresponded to ordinary Bohr radius for a deep impurity
( modified by the fact that in that case one had to consider Dirac Hamiltonian for the
$k-p$ model of semiconductors). In our case, the large length scale comes from
the hybridization of the localized impurity level with conduction and valence bands.
We will also see below in Section VI that effective length scale, at which direct interaction between impurities
starts to matter, grows logarithmically with decreasing  temperature in the framework of percolation theory or
virial expansion. Thus, at low temperatures, such as Curie temperature, the effective range of interaction
between two impurities becomes even larger.

\section{The exchange Hamiltonian.}

\begin{figure}
\includegraphics[width=3in]{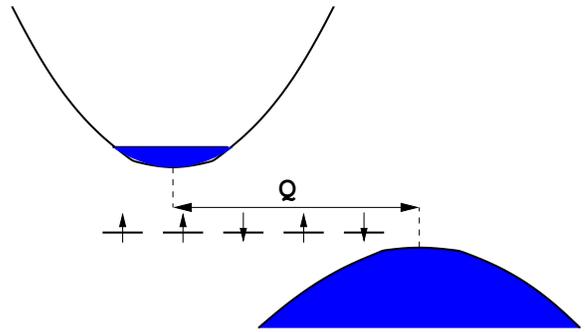}
\caption{Positions of conduction and valence bands.}
\label{bandposition}
\end{figure}

The problem of magnetism in the Anderson model can be reduced to the spin Hamiltonian
Eq.(\ref{exchangeint}), described by the processes shown in Figs \ref{fig4orda}, \ref{fig4ordb}. In case
when the empty band (the "empty" band can also be the valence band, in case of hole doping)
gets partially filled by carriers (electrons or holes), the time-ordered
processes shown in Figs \ref{fig4orda}, \ref{fig4ordb}
include all possible contributions - the familiar RKKY exchange interaction, the superexchange,
and the Bloembergen-Rowland interaction. In presence of carriers, the superexchange gets modified
by the RKKY exchange term. We shall see below, that the Bloembergen-Rowland interaction  also gets
modified in presence of dopants. This modification is not accounted for by the
RKKY interaction. As we have discussed in the previous section, localized impurities in the
Anderson model acquire a large radius (compared with the lattice constant, $a$). Thus, exchange
interaction between localized spins is, in general, long range. This would allow
one to use mean field to describe the ferromagnetic ordering, if the concentration of localized spins
is high enough. On the other hand, if the concentration of magnetic impurities is low, one can
use percolation theory or virial expansion. 

Let us first consider exchange integrals given by the processes shown in
Figs \ref{fig4orda}, \ref{fig4ordb} for a semiconductor with conduction and valence
bands separated by some general reciprocal space wavevector $\bm{Q}$, as shown in Fig.\ref{bandposition}.
When $\bm{Q}=0$, it is a direct band gap semiconductor.
Surprisingly, all exchange processes can be written in a relatively compact way:
\begin{equation}
J(R) = \sum_{i,j} J_{ij}(R),
\label{exchch}
\end{equation}
where
\begin{widetext}
\begin{equation}
J_{ij} = 2 \frac{\left|V_i\right|^2 \left|V_j\right|^2 a^6}{(2 \pi)^6}\,
\int d^3 \bm{p} d^3 \bm{q} \frac{[1 - n(\epsilon_{\bm{p} i} - \mu)] e^{i \bm{q} \bm{R}}}{
(\epsilon_{\bm{p} i} - \epsilon_{\bm{p}+\bm{q} j})(\epsilon_{\bm{p} i} - \epsilon_0)^2}.
\label{exch}
\end{equation}
\end{widetext}

\begin{figure}
\includegraphics[width=3in]{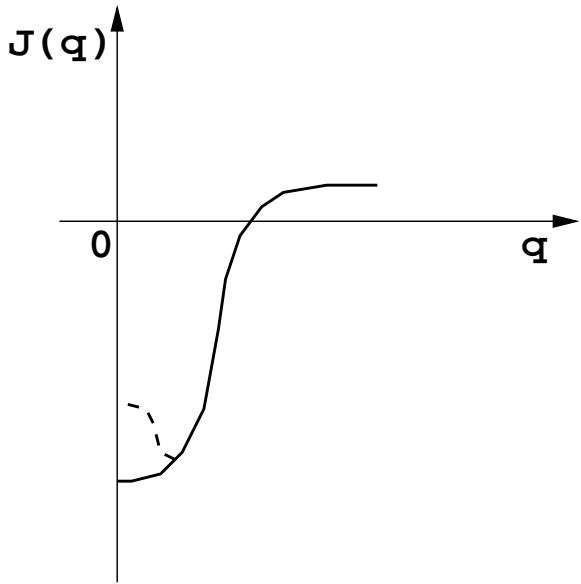}
\caption{Superexchange contribution to exchange integral in q-space; modification by RKKY in presence of carriers
is shown by a dotted line.}
\label{superexchange}
\end{figure}
\begin{figure}
\includegraphics[width=3in]{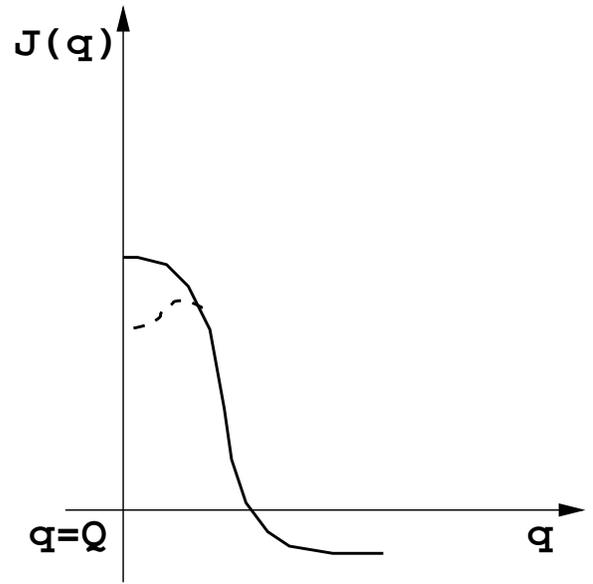}
\caption{Bloembergen-Rowland contribution to exchange integral in q-space; modification in presence of carriers
is shown by a dotted line.}
\label{BlR}
\end{figure}

Here ${i,j}$ are the band indices for conduction and valence bands, $\mu$ is the chemical potential,
$n(\epsilon - \mu)$ is the Fermi-Dirac distribution function. The empty band corresponds to $i=1$,
while the filled band has the band index $i=2$. We can see now, that for $T \ll \Delta_0$, where $\Delta_0$ is the band gap,
the contributions $J_{21}$ and $J_{22}$ are absent. $J_{11}$ is the superexchange, which corresponds to 
the process shown in Fig. \ref{fig4orda}. In presence of carriers, this process, and the corresponding
expression, also includes the RKKY contribution. When carriers are present, the 
effective Hamiltonian can be written in terms of impurity spins only, with carriers "integrated out".
In this case, impurity spins interact with the exchange Hamiltonian Eqs.(\ref{exchch}),(\ref{exch}), which
includes RKKY. If, however, we wish to retain carriers, and the carrier-impurity p-d exchange interaction,
as we described in section II, then the RKKY part will be the part $S'(\beta)$ in the 4-th order
(see Eq.(\ref{subtr})), which has to be subtracted, and the indirect exchange interaction will be
described by Eq.(\ref{exch}) at zero doping, i.e., with the factor $[1 - n(\epsilon_{\bm{p} i} - \mu)]$ in
Eq.(\ref{exch}) replaced by $1$. These two descriptions are only equivalent when the chemical
potential for doped carriers, $\mu \ll \Delta_1$. Basically, when the doped carriers are retained, we "integrate out" electrons
and holes in the original Anderson Hamiltonian up to the scale $\mu$. When we consider the Hamiltonian
for localized spins only, we integrate out all carriers. 
It makes sense to define the RKKY interaction
for the Anderson model as the difference between the doped and undoped cases in Eqs.(\ref{exchch}),(\ref{exch}).
When $\mu \sim \Delta_1$, the Anderson model \textit{cannot} be reduced to the $p-d$ model, except near the Fermi surface. 
The Anderson model expression for the RKKY interaction in real space  
will then \textit{differ} from the corresponding $p-d$ model expression
at \textit{short} distances, which are important for ferromagnetism. We won't consider
this subtle point any further, since we will always assume the carrier density in FS to be small,
i.e., $\mu \ll \Delta_1$.  In this section and the next section we will deal with an
effective Hamiltonian for spins only, i.e., when all carriers are integrated out. This procedure
for the Anderson model is always justified (although, as we commented above, a reduction to the $p-d$ model is \textit{not}).
$J_{12}$ is the Bloembergen-Rowland interaction, which
also gets somewhat modified by doped carriers. Note that the expression for the exchange integral Eq.(\ref{exch}),
obtained from the Anderson model, differs from the original expression of Bloembergen and Rowland \cite{BR};
it includes the term $(\epsilon_{\bm{p} 1} - \epsilon_0)^2$ in the denominator, which makes the 
exchange integral in the Anderson model long range. Both contributions to exchange integral can be evaluated 
numerically and analytically; they are shown schematically in Figs \ref{superexchange} and \ref{BlR}. 

We see that the superexchange is, in general, antiferromagnetic, while the Bloembergen-Rowland 
contribution favors a spin density wave with a wavevector $\bm{Q}$, separating conduction and valence bands.
Both exchange processes are, in general, of the same order, and their relative strength is determined by
the corresponding hybridization parameters $V_1$ and $V_2$. The superexchange integral in the $\bm{q}$ - space,
in the absence of carriers, can be easily found analytically:
\begin{equation}
J_{11}(\bm{q}) = - \frac{2 V_1^4 a^3 m_1^3}{\pi \sqrt{2 m_1 \Delta_1}}\, \frac{1}{q^2 + 8 m_1 \Delta_1}
\label{superex}
\end{equation}
In presence of carriers, $J_{11}(\bm{q})$ gets modifed by the RKKY interaction:
\begin{widetext}
\begin{equation}
\delta J_{11}(\bm{q}) \equiv J_{RKKY}(\bm{q}) =  \frac{V_1^4 a^3 m_1 p_F}{(2 \pi)^2 \Delta_1^2} \,
\left\{ 1 + \left(\frac{p_F}{q}\, - \frac{q}{4 p_F}\, \right) \ln 
\left(\frac{1 + \frac{q}{2 p_F}\,}{1 - \frac{q}{2 p_F}\,}\right)\right\},
\end{equation}
\end{widetext}
where $p_F = \sqrt{2 m_1 \mu}$ is the Fermi momentum for doped carriers. The superexchange and RKKY contributions
can be easily rewritten in real space:
\begin{equation}
J_{11}(R) = J_{se}(R) + J_{RKKY}(R),
\label{j11}
\end{equation}
with
\begin{equation}
J_{se}(R) = - \frac{V_1^4 a^6 m_1^3}{2 \pi^2 \sqrt{2 m_1 \Delta_1} R}\, \exp\{-\sqrt{8 m_1 \Delta_1} R\},
\label{sesese}
\end{equation}
and a standard expression for the RKKY interaction, taking into account that $J_{pd} \equiv 2 V_1^2/\Delta_1$
(note that in Eq.(\ref{exchangeint}) we sum over each pair of impurities twice):
\begin{equation}
J_{RKKY}(R) = - J_{RKKY} F(2 p_F R),
\end{equation}
where
\begin{equation}
F(x) = \frac{\cos(x)}{x^3}\, - \frac{\sin{x}}{x^4},
\end{equation} 
and
\begin{equation}
J_{RKKY} = \frac{V_1^4 a^6 m_1 p_F^4}{\pi^3 \Delta_1^2}.
\end{equation}
We define the RKKY interaction in the Anderson model formulation as the difference between $J_{11}(R)$
in doped and undoped cases, given by Eq.(\ref{exch}). As we emphasized above, since it depends explicitly 
on impurity energy level,  when $\Delta_1 \sim \mu$, this expression differs from the usual RKKY form. 
Its asymptotic behavior at large distances, certainly, does not change.
In what follows we assume that the carrier concentration is always low, $\mu \ll \Delta_i$, so that
$J_{RKKY}(R)$ is given by the standard expression. 
Note that if $J_{pd}$ and the level position are known, the exchange integral Eq.(\ref{j11}) contains
no free parameters:
\begin{widetext}
\begin{equation}
J_{11}(R) =  - J_{RKKY} 
\left(F(2 p_F R) + \frac{\pi (m_1 \Delta_1)^{3/2}}{2 \sqrt{2} p_F^4 R}\, e^{-2 \sqrt{2 m_1 \Delta_1} R} \right).
\label{11ex}
\end{equation}
\end{widetext}
Similar to superexchange, the Bloembergen-Rowland interaction can also be written in two parts. The first
part is from the empty and filled bands, while the second part takes doped carriers into account:
\begin{equation}
J_{12}(\bm{q}) = J_{BR}(\bm{q}) + J'_{12}(\bm{q}).
\label{12ex}
\end{equation}
The first contribution in Eq.(\ref{12ex}) favors a spin density wave with a wavevector $\bm{Q}$, separating
conduction and valence bands, and has a rather cumbersome form:
\begin{widetext}
\begin{equation}
J_{BR}(\bm{q}) =  \frac{a^3 V_1^2 V_2^2 m_1^2 m_2}{\pi \sqrt{2 m_1 \Delta_1}}\,
\left\{\frac{m_1 \Delta_1 + m_2 \Delta_1 + \frac{q^2}{2}\, + m_2 \Delta_0 - \sqrt{2 (m_1 + m_2) \Delta_1}
\sqrt{2 m_2 \Delta_0 + q^2 \frac{m_2}{m_1 + m_2}\,}}{\Delta_1^2(m_1+m_2)^2 + \Delta_1 (q^2 (m_1-m_2) -
2 m_2 (m_1+m_2) \Delta_0) + \left(\frac{q^2}{2}\,+ m_2 \Delta_0 \right)^2}\right\}. 
\label{BRl}
\end{equation}
\end{widetext}
Here and below in this section the notation will be slightly different; $\bm{q}$ is now the \textit{difference} 
from $\bm{Q}$ (the wavevector separating the bottoms of the two bands), i.e., we
change our notation in the following way: $\bm{q - Q} \longrightarrow \bm{q}$. 
In general, the Bloembergen-Rowland contribution Eq.(\ref{BRl}) is quite cumbersome, which makes it impossible to
find an explicit analytical expression
in real space, except in some limiting cases. However, the leading asymptotic behavior of $J_{BR}(R)$ at large distances 
can be derived. If the deep impurity
level lies inside the gap, $\Delta_1 < \Delta_0$, we find an exponential decay of exchange correlations:
\begin{widetext}
\begin{equation}
J_{BR}(\bm{R}) \simeq \frac{a^6 V_1^2 V_2^2 m_1 m_2}{2 \sqrt{2} \pi^2 R}\,\left(\sqrt{\frac{m_1}{\Delta_1}\,} -
\sqrt{\frac{m_2}{\Delta_0 - \Delta_1}\,} \right) \cos{(\bm{Q} \cdot \bm{R})} e^{-R/R_1},
\label{undBR}
\end{equation}
\end{widetext}
where the range of the integral is:
\begin{equation}
R_1 = \frac{1}{\sqrt{2 m_1 \Delta_1} + \sqrt{2 m_2 (\Delta_0 - \Delta_1)}}.
\label{r1}
\end{equation}
On the other hand, when the deep impurity level enters the filled band, $\Delta_1 > \Delta_0$,  the Bloembergen-Rowland 
contribution has an oscillating decaying asymptotic:
\begin{widetext}
\begin{equation}
J_{BR}(\bm{R}) = \frac{a^6 V_1^2 V_2^2 m_1 m_2}{ \sqrt{2} \pi^2 R}\, \left(\sqrt{\frac{m_1}{2 \Delta_1}\,} \cos\frac{R}{R_{c1}}\, +
\sqrt{\frac{m_2}{2 (\Delta_1-\Delta_0)}\,} \sin\frac{R}{R_{c1}}\,\right) \cos{(\bm{Q} \cdot \bm{R})} e^{- R/R_{c0}},
\end{equation}
\end{widetext}
where
\begin{equation}
R_{c1} = \frac{1}{\sqrt{2 m_1 (\Delta_1 - \Delta_0)}}, \ \ \ R_{c0} = \frac{1}{\sqrt{2 m_1 \Delta_1}}\, .
\label{rc1rc0}
\end{equation}

The contribution from doped carriers to $J_{12}$ can be derived as well:
\begin{equation}
J'_{12}(\bm{q}) \simeq - \frac{ a^3 V_1^2 V_2^2 p_F^3}{3 \pi^2 \Delta_1^2 (\Delta_0 + \frac{q^2}{2 m_2}\,)}\,
\end{equation} 
In real space, this contribution becomes:
\begin{equation}
J'_{12}(\bm{R}) \simeq - \frac{p_F^3 a^6 V_1^2 V_2^2 m_2}{6 \pi^3 \Delta_1^2 R} \cos{(\bm{Q} \cdot \bm{R})} e^{-\sqrt{8 m_2 \Delta_0} R}
\label{dopBR}
\end{equation}

We see that if $\bm{Q}=0$, i.e., when our FS has a direct band gap, the Bloembergen-Rowland mechanism gives
a large ferromagnetic short-range contribution to the exchange integral.
At high enough concentration of impurities, ferromagnetic properties are then determimed by the value of exchange
integral at $\bm{q}=0$. This value can be easily written down (it is only valid for a direct band gap FS):

\begin{widetext}
\begin{equation}
J(\bm{q}=0) = J_{11}(\bm{q}=0) + J_{12}(\bm{q}=0)
 = \frac{2 a^3 V_1^2 m_1^2}{ \pi \sqrt{2 m_1 \Delta_1}}\, \left(- \frac{V_1^2}{8 \Delta_1}\, 
+ \frac{V_2^2 m_2}{ (\sqrt{2 (m_1 + m_2) \Delta_1} + \sqrt{ 2 m_2 \Delta_0})^2}\, \right) 
\label{qzero}
\end{equation}
\end{widetext}

In this section we have derived explicitly the interaction between two magnetic impurities in a semiconductor
with one conduction and one valence band. Typically, the band structure of semiconductors is more complex
than that (for example, GaAs has light and heavy hole bands). 
This situation is considered in the next section.

\section{What if a semiconductor has more than two bands?}

When a semiconductor has many bands, the analysis is just as straightforward as in case
of two bands considered in the previous section. The exchange integral between two impurities
is still given by Eqs(\ref{exchch}),(\ref{exch}), where the sum now goes over all pairs of band indices. The case when
$i=j$ is an ``empty'' band index corresponds to a superexchange contribution
(Fig.\ref{fig4orda}), considered in detail in the previous section; $J_{ij} = 0$ when $i$ is a ``filled''
band index, because of the Pauli principle; $J_{ij}$ is the Bloembergen-Rowland contribution (Fig.\ref{fig4ordb}), 
when $i$ is an ``empty''
band index, and  $j$ is a ``filled'' band index. The contribution of a new type appears when one has
two or more ``empty'' bands in a semiconductor. Then there will be a superexchange contribution of the type shown in
Fig.\ref{fig4orda}, where now the carriers are exchanged through two different ``empty'' bands - the process (1) in
Fig.\ref{fig4orda} puts a carrier from one magnetic impurity into the first ``empty'' band, while the process (2)
puts a carrier from the other magnetic impurity into the second ``empty'' band. Naturally,
there are two such contributions, $J_{ij}$ and $J_{ji}$, where $i \neq j$ indices correspond to two
different ``empty'' bands. The exchange Hamiltonian in any FS
is a sum of all pairwise contributions listed above. 

Let us now consider the new two-band superexchange contribution in detail.
In the most general case, 
the bottom of the second ``empty'' band is shifted from the bottom of the first ``empty'' band by a 
wavevector $\bm{Q}$. The bottoms of the two bands also lie at two different positive
energies $\Delta_1$ and $\Delta_2$ relative to impurity level, unless there is a symmetry-related 
degeneracy, and the carriers in the first and second bands have different 
effective masses $m_1$ and $m_2$. 

  Let us first consider the simplified case when the bands 1 and 2 are identical (i.e., $\Delta_1 = \Delta_2$ and
$m_1 = m_2$), but separated by a  wavevector $\bm{Q}$. Then we don't have to do the calculation, since  
$\epsilon_{\bm{p}2} = \epsilon_{\bm{p-Q} 1}$, and it can
be easily seen from Eq.(\ref{exch}), that
\begin{equation}
J_{12}(\bm{q}) = J_{11}(\bm{q-Q}),
\end{equation}
and
\begin{equation}
J_{12}(R) = J_{11}(R) e^{i \bm{Q} \bm{R}}.
\end{equation}
Summing all contributions from 2 identical empty bands, since $J_{11}(R) = J_{22}(R)$,
we get:
\begin{equation}
J_{2bse} = 2 J_{11}(R) [1 + \cos{(\bm{Q} \cdot \bm{R})}].
\label{2bse}
\end{equation}
This also includes RKKY-type contribution for two identical bands at finite doping, described by:
\begin{equation}
J_{12 RKKY} + J_{21 RKKY}  = 2 J_{RKKY}(R) \cos{(\bm{Q} \cdot \bm{R})},
\label{2bse1}
\end{equation}
and the ordinary RKKY interaction from both bands.
It is rather obvious from Eq.(\ref{2bse1}), that at a finite density of carriers there are ordinary one-band RKKY contributions from
both bands. The two-band RKKY contribution oscilates much more rapidly in space, 
as $\cos(\bm{Q}{R}))$, unless $\bm{Q = 0}$. In the latter case, the two-band RKKY contribution just enhances
the contributions from the two separate bands. In the case when $\bm{Q}=0$, the $12$ and $21$ exchange integrals take
the same form as shown in Fig.\ref{superexchange}. For $\bm{Q \neq 0}$, these contributions have the
same form in $\bm{q}$-space as for $\bm{Q}=0$, centered at the wavevector $\bm{q}=\bm{Q}$.

  Now that the form of new two-band contributions to the exchange integral has become clear from the simplified case, 
let us consider a more 
general case of two completely different ``empty'' bands. The bottom of the second band is shifted from
the bottom of the first band by the wavevector  $\bm{Q}$. Of course, $12$ and $21$ contributions to $J(\bm{q})$
will still be of the form shown in Fig.\ref{superexchange}, with the bottom at $\bm{q} = \bm{Q}$, but the corresponding 
expressions become more combersome. Let us assume that $\Delta_1 < \Delta_2$, i.e. the bottom of the first band
lies below the bottom of the second band, and the first band gets filled
by carriers first. We consider three separate cases, (1) both bands are empty; (2) the first band gets
filled by carriers, but the second band is empty, and (3) both bands get partially filled by carriers.
When $\Delta_1 = \Delta_2$, i.e., the two bands are symmetry-related, we only have cases (1) and (3). 
As before, we assume that the carrier
concentration is very small, i.e., the Fermi energy for doped carriers is much smaller than any other 
energy scale in the problem, except temperature.

\vglue 0.5cm

\noindent
\textit{(1) Both bands are empty.}

\vglue 0.5cm

  Let us now consider the exchange interaction arising from two empty bands. The integral 
in Eq.(\ref{exch}) can be easily calculated:
\begin{widetext}
\begin{equation}
J_{12+21}(\bm{q}) = - \frac{\sqrt{2} V_1^2 V_2^2 m_1 m_2 a^3}{(\bm{q}-\bm{Q})^2 + 
(\sqrt{2 m_1 \Delta_1} + \sqrt{2 m_2 \Delta_2})^2}\, \left(\sqrt{\frac{m_1}{\Delta_1}\,} 
+\sqrt{\frac{m_2}{\Delta_2}\,}\right).
\end{equation}
\end{widetext}
In the coordinate space, it takes the following form:
\begin{equation}
J_{12+21}(\bm{R}) = - V_{12} \cos{(\bm{Q}\cdot\bm{R})} \frac{1}{R} \exp{( - \frac{R}{R_{12}}\, )},
\label{2eb}
\end{equation}
where the range of this interaction is given by:
\begin{equation}
R_{12} = \frac{1}{\sqrt{2 m_1 \Delta_1} + \sqrt{2 m_2 \Delta_2}}, 
\end{equation}
and
\begin{equation}
V_{12} = \frac{V_1^2 V_2^2 a^6 m_1 m_2}{2 \sqrt{2} \pi^2}\, \left(\sqrt{\frac{m_1}{\Delta_1}\,} + 
 \sqrt{\frac{m_2}{\Delta_2}\,} \right)
\end{equation}

\vglue 0.5cm

\noindent
\textit{(2) One band gets filled.}

\vglue 0.5cm

Let us assume in this section that band one gets filled first, i.e., $\Delta_1 < \Delta_2$.
We also assume that the chemical potential (counted from the bottom of band one), $\mu_1 \ll \Delta_2 - \Delta_1$,
i.e., the number of carriers is low. Of course, the main part of the exchange inegral is still given by Eq.(\ref{2eb}).
However, at finite doping there are corrections. Obviously, since only band one gets carriers, $\delta J_{21} = 0$.
At small filling, RKKY-like correction can be easily calculated:
\begin{equation}
\delta J_{12}(q) \simeq \frac{2 V_1^2 V_2^2 a^3 P_{F1}^3 m_2}{3 \pi^2 \Delta_1^2 (2 m_2 [\Delta_2 - \Delta_1] + q^2)}.
 \end{equation} 
Here $q$ denotes the difference from $\bm{Q}$, the wavevector separating the bottoms of the two bands.
In coordinate space it can be written as:
\begin{equation}
\delta J_{12}(\bm{R}) \simeq
\frac{V_1^2 V_2^2 a^6 p_{F1}^3 m_2}{6 \pi^3 \Delta_1^2 R} e^{- R/R_g} \cos{\bm{Q} \cdot \bm{R}},
\end{equation} 
where
\begin{equation}
R_g = \frac{1}{\sqrt{2 m_2 (\Delta_2 - \Delta_1)}}\,
\end{equation}

\vglue 0.5cm

\noindent
\textit{(3) Both bands get filled.} 

\vglue 0.5cm

When $\Delta_1 = \Delta_2 = \Delta$ by symmetry, both bands get filled by carriers simultaneously.
The result of this is an RKKY-like correction to exchange integral  Eq.(\ref{2eb}), which is given
by Eq.(\ref{2bse1}) for two identical bands. If the bands have different masses, the interaction
is RKKY-like long range, but the general expression is quite messy. Here we give the its value at 
$\bm{q}=\bm{Q}$, which is important for ferromagnetism, when $\bm{Q} = 0$:

\begin{equation}
 \delta J_{12+21}(\bm{q}=\bm{Q}) = \frac{2 V_1^2 V_2^2 a^3 \sqrt{2 \mu} m_1 m_2}{\pi^2 \Delta^2 (\sqrt{m_1} + \sqrt{m_2})}\,
\end{equation}
 
To summarize this section, the exchange interaction in any magnetic semiconductor is composed of pairwise contributions:
the superexchange contribution through one ``empty'' band, the Bloembergen-Rowland contribution from one ``empty'' and one filled band,
and a pairwise contribution from two different ``empty'' bands. The last contribution was considered in this section. 
It includes two-band long-range RKKY-like interaction at finite doping, which is usually not taken into account in
the $p-d$ Hamiltonian. Note that when $\bm{Q} \neq 0$, the two-band RKKY interaction will favor spin glass order,
\textit{not} ferromagnetism, since it oscillates very rapidly in real space.

\section{The mean field approximation, virial expansion, and percolation.}

   Now that we have obtained all terms in the exchange interaction Eq.(\ref{exch}), we can proceed to
calculate magnetic properties of ferromagnetic semiconductors. In general, the RKKY exchange interaction
is always long-range. The shorter-range contributions may or may not be treated in mean field, depending
on the ratio of their range to the average inter-impurity distance. We assume here, for simplicity,
that one of indirect exchange processes considered in the previous two sections dominates the short-range
physics. In case of FS with indirect band gap, the Bloembergen-Rowland exchange contribution favors 
antiferromagnetism with a lattice wavevector $\bm{Q}$. It oscillates very rapidly in real space. 
Since impurities are distributed randomly,  it would favor a spin glass ordering.
 The contributions at $\bm{Q}=0$, which influence ferromagnetism, are the RKKY
and the superexchange. From Eq.(\ref{11ex}) one can easily see that, when impurity concentration
$n_i \gg (m_1 \Delta_1)^{3/2}$, the superexchange contribution totally suppresses ferromagnetism 
brought about by the
RKKY exchange interaction. On the other hand, when $n_i \ll (m_1 \Delta_1)^{3/2}$, the superexchange
at an average distance between impurities is
suppressed, and the RKKY interaction gives rise to ferromagnetism, as in the ordinary $pd$-model. Corrections
due to superexchange can then be calculated using the virial expansion approach\cite{BA}. We can
rewrite the dominant exchange contribution from Eq.(\ref{11ex}) in the following form:
\begin{equation}
J(R) = - J_{RKKY} F(2 p_F R) - V_0 \frac{R_0}{R}\, e^{-R/R_0},
\label{generic}
\end{equation}
where $V_0 = \pi J_{RKKY}/(64 p_F^4 R_0^4)$, $R_0 = 1/\sqrt{8 m_1 \Delta_1}$, in case when it is given by one empty band only.
If contributions from more than one ``empty'' band are important, $V_0$ and $J_{RKKY}$ are, in general, unrelated to each
other. The reason for this is that the largest short range contribution typically comes from the lightest bands, 
which would produce the longest range indirect exchange (if all hybridization parameters
are of the same order). On the other hand, the dominant
RKKY contribution is the one from the heaviest band, since the carriers in that band have the largest Fermi wavevector; there
will also be additional RKKY contributions from many bands, which were analyzed in the previous section.
In what follows we consider the simplest case, when only one heavy band is relevant for RKKY. 

In case of a FS with a direct band gap, the Bloembergen-Rowland mechanism gives rise to ferromagnetism, and
we can either have a ferromagnetic or an antiferromagnetic short-range exchange contribution, depending
on the relative strength of corresponding exchange processes. If the Boembergen-Rowland term
dominates the physics at short distances, the impurity spin Hamiltonian becomes:
\begin{equation}
J(R) = - J_{RKKY} F(2 p_F R) + V_{BR} \frac{R_1}{R}\, e^{-R/R_1},
\end{equation}
In general, the Bloembergen-Rowland term Eq.(\ref{12ex}) also has an antiferromagnetic contribution from
doped carriers. It then can be rewritten in the following form:
\begin{equation}
J_{12}(R) = V_{BR} \frac{R_1}{R} e^{-R/R_1} - V_{dop} \frac{R_g}{R_0} e^{-R/R_g},
\end{equation}
where $R_g = 1/\sqrt{8 m_2 \Delta_0}$, $V_{BR}$ and $V_{dop}$ are given by the coefficients
in front of exponents in Eq.(\ref{undBR}) and Eq.(\ref{dopBR}). However, since
\begin{equation}
V_{dop}/V_{BR} \sim \frac{p_F^3}{(m_1 \Delta_1)^{3/2}} \ll 1,
\end{equation}
while $R_g$ is still quite small, the contribution of doped carriers to $T_c$ through the Bloembergen-Rowland mechanism 
can be neglected. Thus, if one exchange process dominates the physics are short distances, the short-range 
exchange contribution has the same form, and can only differ in sign.
In what follows, we consider the general form of exchange integral Eq.(\ref{generic}), assuming that the exchange
constant $V_0$, its sign, and its range $R_0$ are those of the dominant indirect exchange contribution.
Note again that in our definition of exchange integral we sum over impurities twice.
We will also consider the case when the short-range part in Eq.(\ref{generic}) is rapidly oscillating,
\begin{equation} 
 J_{SR} = - V_0 \frac{R_0}{R}\, e^{-R/R_0} \cos{(\bm{Q}\cdot\bm{R})}.
\label{Osc}
\end{equation}
As we have seen above, this happens when  two bands separated by wavevector $\bm{Q}$ contribute
the most to the short-range exchange interaction. $V_0$ has a negative sign for the Bloembergen-Rowland
contribution, and a positive sign for the superexchange.
When more than one process is important for short-range physics,
the problem can always be solved numerically for a given set of parameters.
Here we obtain an analytical solution in several limiting cases.

\vglue 0.5cm

\noindent
\textit{(1) Dilute system with almost no carriers.}

\vglue 0.5cm

In the absence of carriers, the type of order and $T_c$ is determined by the short-range part of 
interaction. If ferromagnetic Bloembergen-Rowland interaction dominates at short distances, Curie temperature
is approximately given by the ferromagnetic interaction taken at the average distance between impurities
\cite{Korenblit}:
\begin{equation}
T_c \simeq  2.3 V_0 S^2 R_0 n_i^{1/3} \, e^{- \frac{0.87}{R_0 n_i^{1/3}}} \ \ V_0 > 0,
\label{perc}
\end{equation}
which is valid when $n_i \ll 1/R_0^3$. A comparison with Eq.(\ref{Curie}) shows that this
gives the following condition on the number of carriers:
\begin{eqnarray}
\label{cond}
p_F &=& (3 \pi^3 n_e)^{1/3} \ll p_{0F} \\
p_{0F}&=&\frac{27.6 V_0 \pi^2 S R_0}{(J^{pd}_{\bm{q}=0})^2 (S+1) n_i^{2/3}}\, e^{-0.87/(R_0 n_i^{1/3})} \nonumber
\end{eqnarray} 
When the short-range interaction is antiferromagnetic, it favors spin glass order, and 
ferromagnetism is absent when there are no or almost no carriers.
The condition on the number of carriers for ferromagnetism to be absent is then given by:
\begin{equation}
p_F \ll p_{0SG}=\frac{27.6 |V_0| \pi^2 R_0}{(J^{pd}_{\bm{q}=0})^2 n_i^{2/3}}\, e^{-0.87/(R_0 n_i^{1/3})}
\label{cond1}
\end{equation}
Finally, for a rapidly oscillating short-range part Eq.(\ref{Osc}) this condition can be rewritten as:
\begin{equation}
p_F \ll p_{0SG}=\frac{27.6 |V_0| \pi^2 R_0}{(S+1)(J^{pd}_{\bm{q}=0})^2 n_i^{2/3}}\, e^{-0.87/(R_0 n_i^{1/3})}
\label{cond2}
\end{equation}

For the Anderson Hamiltonian, $J^{pd}_{q=0} = 2 V_1^2 a^3/\Delta_1$. Note that if the ratio $n_e/n_i$ is fixed,
$p_F$ grows much slower with $n_i$ than $p_0$. Thus, conditions in  Eqs.(\ref{cond}),(\ref{cond1}),(\ref{cond2})
are likely to be satisfied at some finite concentration of magnetic impurities, $n_c \ll  n_i \ll 1/R_0^3$.
For example, for antiferromagnetic or oscillating interactions, Eqs (\ref{cond1}),(\ref{cond2}) will define the concentation 
of impurities above which carrier-driven ferromagnetism disappears. 

\vglue 0.5cm

\noindent
\textit{(2) Dilute system with carriers.}

\vglue 0.5cm

When the carrier concentration is large, i.e., $p_F \gg p_0$ in Eqs.(\ref{cond}),(\ref{cond1}),(\ref{cond2}),
but the concentration of magnetic impurities is still small, $n_i \ll 1/R_0^3$, short-range interactions 
between magnetic impurities will result in a correction to $T_c$, which
can be calculated by virial expansion (see, for example, Ref.\cite{BA}). Following Ref.\cite{AG},
since the range of RKKY interaction is large, we can represent $p-d$ interaction between carriers 
and magnetic impurities by a mean field Zener Hamiltonian:
\begin{equation}
\hat{H}_{MF} =  - J^{pd}_{\bm{q}=0} \bm{s} \sum_i \bm{S}_i, 
\label{zener}
\end{equation}
where $\bm{s}$ is the density of ordered spin of the carriers, which is assumed to be constant
in space. In addition, there is a relatively short-range exchange interaction between impurity spins,
given by the processes described in Sections IV and V: 
\begin{equation}
\hat{H}_{exch} = - \sum_{ij} J(\bm{R}_i - \bm{R}_j) \bm{S}_i \bm{S}_j.
\label{exx}
\end{equation}
The short-range exchange integral does not include the carrier contribution, since it is already 
accounted for in the mean field Hamiltonian.
In the Zener model Eq.(\ref{zener}) , Curie temperature Eq.(\ref{Curie}) can be found by minimizing the free energy 
density of the system of carriers and spins\cite{AG},
\begin{equation}
F = F_e + F_i,
\end{equation}
with respect to $\bm{s}$, and finding when the solution at small $s$ first appears. Here
\begin{equation}
F_e= \frac{(2 \mu_B s)^2}{2 \chi_0},
\end{equation}
and $F_i$ is given by the usual Zeeman term,
\begin{equation}
F_i = - n_i T \ln{\frac{\sinh{[J^{pd}_{\bm{q}=0} s (S + 1/2)/T]}}{\sinh{[J^{pd}_{\bm{q}=0} s/(2 T)]}}\,} 
\end{equation}

To find the virial correction to Eq.(\ref{Curie}) from Eq.(\ref{exx}), we need to include into $F$ the
contribution from two magnetic impurities, when they are close enough:
\begin{equation}
F_{2i} = \frac{1}{2}\, n_i^2 \int d^3 \bm{R} [F_{2i}(\bm{R}) - 2 F_i (\bm{R})],
\end{equation}
and calculate it from the Hamiltonian
\begin{equation}
\hat{H} = \hat{H}_{MF} + \hat{H}_{exch}.
\end{equation}
This can be done, since the Zeeman Hamiltonian for two spins, interacting via direct
exchange interaction, can easily be solved. The integral over this solution, however,
can only be taken with logarithmic accuracy. The contribution from two impurities is important
when the distance between them is
\begin{equation}
R \le R_0 \ln{\frac{|V_0|/T}{\ln{(|V_0|/T)}}}.
\end{equation}
Finding $F_{2i}$, and repeating the minimization over $s$, we obtain, as expected, that a for ferromagnetic exchange
interaction ($V_0 > 0$) Curie temperature is enhanced:
\begin{equation}
\frac{\delta T_c[F]}{T_c} \simeq \frac{4 \pi S}{3 (S + 1)}\, n_i R_0^3 \ln^3{\frac{|V_0|/T_c}{\ln{(|V_0|/T_c)}}}.
\label{ferr}
\end{equation}
For an antiferromagnetic exchange interaction ($V_0 < 0$), the Curie temperature is reduced:
\begin{equation}
\frac{\delta T_c[A]}{T_c} \simeq - \frac{4 \pi}{3}\, n_i R_0^3 \ln^3{\frac{|V_0|/T_c}{\ln{(|V_0|/T_c)}}}.
\label{aferr}
\end{equation}
A rapidly oscillating exchange interaction Eq(\ref{Osc}) gives:
\begin{equation}
\frac{\delta T_c[O]}{T_c} \simeq - \frac{4 \pi}{3 (S+1)}\, n_i R_0^3 \ln^3{\frac{|V_0|/T_c}{\ln{(|V_0|/T_c)}}},
\label{oscill}
\end{equation}
which is independent of the sign of $V_0$.
\vglue 0.5cm

\noindent
\textit{(3) ``Dense'' system.}

\vglue 0.5cm

The most interesting situation is the case of a ``dense'' system of magnetic impurities,
when  $n_i \gg (m_1 \Delta_1)^{3/2}$. Since the short-range part of the interaction could
have a range much larger than the lattice spacing, it need not be really dense.  
This requirement can be rewritten as $n_i \gg (\Delta_1 / D)^{3/2}$. In a dense system,
the wave functions of the neighboring impurities overlap strongly, and the main exchange contribution
arises as a result of this overlap, not the RKKY interaction through free carriers. Then
ferromagnetism arises even when no carriers are present, if the Bloembergen-Rowland exchange process 
dominates the physics at short distances ($V_0>0$):
\begin{equation}
T_c = \frac{2 S (S+1) n_i}{3}\, J_{\bm{q}=0} = \frac{8 \pi S (S+1) n_i}{3} V_0 R_0^3.
\end{equation}
Note that $T_c$ does not depend on the carrier
concentration, and should be much higher than that resulting from the RKKY
interaction. 

The Bloembergen-Rowland exchange integral may be weaker than superexchange. In that case,
ferromagnetism in a ``dense'' system is suppressed, and spin glass order is favored. This is always 
the case for indirect band gap 
semiconductors, where the short-range ferromagnetic exchange is absent.
 
The Bloembergen-Rowland mechanism in case of direct band gap 
necessarily leads to an increase of maximum $T_c$ as a
function of concentration of magnetic impurities $n_i$. 

\section{Application to GaAs:Mn.}

Application of the Anderson model to a real system, such as Ga$_{1-x}$Mn$_x$As is somewhat more 
involved than the model that we considered above, since Mn ion is in 3d$^5$ configuration with a 
spin $S=5/2$. This configuration has 5 d-orbitals. For symmetry reasons, 
there may be more than just one conduction or valence band,
which is the case for GaMnAs. Then, as we discussed above in Section V, one has to take into account all pairwise
contributions to the exchange integral, Eqs.(\ref{exchch}),(\ref{exch}). In particular, 
there may be unusual contributions to RKKY,
such as those considered in Section V.
In general, one has also to sum over all orbitals in the Anderson Hamiltonian, not just spins,
and take the Hund's rule, spin orbit, and crystal field splitting into account. Different orbitals 
may have different $V-s$ with conduction and
valence bands. The result of this treatment, however, produces the same exchange integrals $J(R)$; for
example, in case when spin-orbit and crystal field splitting is neglected, the result will still be
given by the exchange integral Eqs.(\ref{exchch}),(\ref{exch}), with 
$|V_1|^2$ is replaced by $\sum_m |V_{1m}|^2$, and similarly for $|V_2|^2$ ,where $V_{1m}$ is
$V_1$ for $m$-th orbital. The relation between corrections to energy levels and exchange integrals
is different in this more realistic case. However, the integrals involved are the same, and, as
we have seen above in section III, corrections to energy levels are rather small. Taking Hund's rule into 
account results in replacing $S=1/2$ operator
in equations of the previous section by $\bm{S}/(2S)$, with $S = 5/2$. Thus, this leads to
the same results as in the previous section, with somewhat redefined $V-s$. The relationship
between the energy shift and $J_{pd}$ will change, but the
actual form of $J(R)$ is determined by the energy spectrum only. Instead of $V_{\alpha m}$, we may introduce 
\begin{equation}
|V_{\alpha}|^2 \equiv \frac{\sum_m |V_{\alpha m}|^2}{4 S^2}\, ,
\end{equation}
and use the form of exchange integrals that we have obtained in the previous sections (with $S=5/2$).
Equivalently, this would mean that the expression for $J(\bm{R})$ in terms of $J_{pd}$-s (or $J_{RKKY}$) for conduction 
and valence bands and level position will stay the same as in the previous sections.
 
For a $p$-type semiconductor, such as Ga$_{1-x}$Mn$_x$As, we can adopt an inverted picture, 
where the ``empty'' band is now the valence band (empty of holes), while the filled band is the conduction band. 
The Anderson Hamiltonian is then easily rewritten in terms of holes. There are two types of holes in GaAs - 
the heavy hole and the light hole. Since their masses are very different ($0.081 m_e$ and $0.51 m_e$),
the main RKKY exchange contribution is produced by the heavy hole. On the other hand, the superexchange contribution
from the light hole band has a much larger range, and thus could potentially be more important than the superexchange contribution
from the heavy hole band, or the mixed superexchange contribution. However, as we have seen in previous sections, most short range
contributions for a direct band gap semiconductor take the form of the second term in  Eq.(\ref{generic}), although
there may be some variations. The problem is that the amplitude of superexchange $V_0 \propto V^4/D^3 \propto m^3 $, where $D$ is
the band width! So, for light bands the effective range of the interaction is large, but the payback is that the amplitude turns 
out to be small.  An easy estimate for the light hole band in GaAs shows that for $V$-s of the order of $1 eV$ this band plays
no role in ferromagnetism. The hole mass in the split-off band
is $0.15 m_e$. This band could also play an important role in the superexchange interaction, although, once again, the
amplitude for realistic parameters turns out to be extremely small. The electron mass is $m_{ge} = 0.063 m_e$
for the main $\Gamma$-valley. The masses and gaps for $L$- and $X$- valleys are much larger, so we don't expect them
to play much role. Thus, we arrive at a simplified picture, where only the
heavy hole band and $\Gamma$-valley electrons are relevant.
The CFR $Mn$ $d^6/d^5$ level,
which is important for our analysis, is in the conduction band, $\Delta_1 =1.5 eV$ above the top of the valence band. 
We can see from Eqs (\ref{rc1rc0}),(\ref{sesese}) that, for this particular level position, the ferromagnetic  Bloembergen-Rowland 
interaction has the range $R_0 \equiv R_{BR} \simeq \hbar/\sqrt{2 m_{hh} \Delta_1}$, while the range of the superexchange is
$R_{se} \simeq R_{BR}/2$. The amplitude of the Bloembergen-Rowland term, $V_{BR}/V_{se} \simeq V_2^2 m_{ge}/(2 V_1^2 m_{hh})$,
could become comparable or exceed the amplitude of heavy hole superexchange.
Note that, in case of strong short-range ferromagnetic
interaction, it would be energetically favorable for $Mn$ impurities to form ferromagnetic clusters.
This, in turn, would reduce the Curie temperature. Clustering of $Mn$ impurities would make magnetic properties
of this material crucially dependent on sample preparation. 
On the other hand, antiferromagnetic short-range interactions should be stronger at shorter distances, 
which would potentially lead to an exchange integral (in the absence
of carriers), which changes its sign as a function of the distance between impurities. In general, for the particular
situation when the $Mn$ level is almost at the bottom of conduction band,  the range of antiferromagnetic superexchange 
is approximately $R_0/2$, and we may represent the total short-range exchange integral in the following form:
\begin{equation}
J(R) \simeq V_0 \frac{R_0}{R}\, [- \alpha \exp{(- 2 R/R_0)} + \exp{(- R/R_0)}].
\label{GaAsex}
\end{equation}
Here $V_0 = V_{BR} > 0$, while $\alpha \simeq m_{hh} V_1^2/(m_{ge} V_2^2)$ is the ratio of superexchange and Bloembergen-Rowland 
amplitudes (up to a factor of 2), which depends on the hybridization of  
impurity d-level with the valence band $(V_2)$, of impurity d-level with the heavy hole band, $V_{hh}$, and the corresponding effective
masses. When $\alpha > 1$ (which is likely the case here, since $m_{hh} \gg m_{ge}$), the exchange becomes 
antiferromagnetic at short distances for $R < R_0 \ln{\alpha}$. The virial correction to $T_c$
for such exchange integral (assuming $T_c$ is determined mostly by the $p-d$ interaction of magnetic impurities and heavy holes)
is then given by: 
\begin{widetext}
\begin{equation}
\frac{\delta T_c}{T_c} \simeq \frac{4 \pi S}{3 (S + 1)}\, n_i R_0^3 (\ln^3{\frac{V_0/T_c}{\ln{(V_0/T_c)}}} - [2 + (1/S)]\ln^3{\alpha}),
\end{equation}
\label{deltatc}
\end{widetext}
and could change sign as well, at some large doping level. If carriers are not present, this exchange integral alone, for $\alpha > 1$,
would lead to a saturation or decrease of $T_c$ at large doping. On the other hand, when $\alpha < 1$, ferromagnetism
gets significantly enhanced at short distances.

In general, the interplay between various short range contributions leads to a rather complicated physics
at short distances. While a detailed calculation requires precise knowledge of all hybridization parameters
from the quantum chemistry, we can estimate the $Mn$ concentration at which the short distance
physics becomes important by requiring $\delta T_c / T_c \sim 0.5$ in Eq.(\ref{deltatc})
We take the estimate of $J_{pd} \sim 150 eV \AA^3$ and $T_c \sim 110K$ from Ref.\cite{Ohno},  $\Delta_1 \sim 1.5 eV$, 
and assume that $\alpha \simeq 1$. Then $V_0 = J_{pd}^2 m_{hh}^3/(8 \pi^2)$, and we get:
\begin{equation}
x_i \sim \left(\frac{a}{2 R_0 \ln{(V_0/T_c)}}\, \right)^3 \sim 8\%
\end{equation}

\section{Effects of disorder and interactions.}

\begin{figure}
\includegraphics[width=3in]{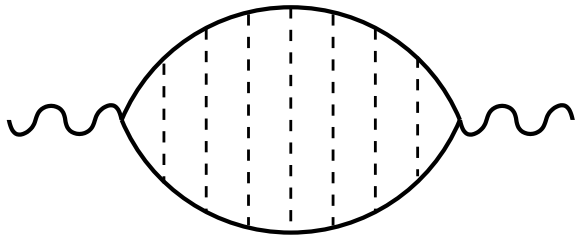}
\caption{Impurity corrections to Curie temperature.}
\label{fig7}
\end{figure}

In this section we consider rather briefly effects of disorder and interactions. Since the superexchange and the Bloembergen-Rowland
exchange interaction are governed by high-energy virtual processes, they are independent of disorder. The RKKY interaction,
however, gets modified. This modification was first considered by de Gennes\cite{deGennes}, who argued that, since the
RKKY interaction at large distances is dominated by the $2 p_F$ Kohn anomaly wavevector, the vertex corrections are
not essential for the averages over disorder. 
The long-distance power law in the RKKY interaction then gets multiplied by an exponential factor $\exp{(- 2 R/l)}$,
where $l$ is the scattering length. These effects were indeed taken into account by Ohno et al.\cite{Ohno} in their
original paper. Abrahams \textit{et al.}\cite{elihu}, however, have shown, that this is not the whole story, 
since disorder introduces instead
a \textit{distribution} of $J(R)$ at large distances. We note here that the long-distance behavior
of the RKKY interaction is not
essential for ferromagnetism. The Curie temperature is determined by RKKY exchange at \textit{short} distances,
or $J_{RKKY}(\bm{q}=0)$. Of course, the vertex corrections are essential for the calculation of the RKKY loop diagram at
$\bm{q} = 0$. Summing all ladder diagrams ($0$ order in $1/p_F l$),  shown in Fig. \ref{fig7}, leads to a diffuson contribution, 
\begin{equation}
\Pi(\bm{q}, \omega_n) = - \frac{\nu D \bm{q}^2}{|\omega_n| +  D \bm{q}^2}\, ,
\end{equation}
which significantly modifies frequency dependence of RKKY at $\bm{q} = 0$. Here $D$ is the diffusion coefficient.  

However, the static ($\omega = 0$), not dynamic, part of the diagram in Fig. \ref{fig7}  
determines $T_c$, and it is not changed at all. 
Thus, to the leading order in $1/p_F l$ disorder does not modify the Curie temperature. 
The interactions, if not too strong, can also be included as the standard Fermi-liquid corrections to $\chi_0$ in Eq.(\ref{Curie}).
Weak localization corrections (the Cooperon diagrams), however, should modify $T_c$. They can also be included in
the same way as the standard weak localization corrections to spin susceptibility (see, for example, Ref.\cite{lee}), 
and Eq.(\ref{Curie}) should still be valid. 

Finally, we note that strong exchange interaction $J^{pd}$ could bind holes at $Mn$ sites, forming a shallow
(or deep) complex magnetic impurity. This effect would
reduce the hole concentration and the number of free $Mn$ spins, and thus lead to a reduction Curie temperature. 
The interactions between
these complex magnetic impurities would be determined by the overlap of the corresponding wavefunctions.  

\section{Conclusions.}

   We have investigated the model of magnetic semiconductors in which magnetic impurities are treated in the
framework of the Anderson model. We have shown that the effective Hamiltonian of this model is more rich than
the usual p-d model considered in the literature. Effectively, in the Anderson model, wave functions of localized
impurities develop a "tail", which could be long range. When the concentration of magnetic impurities is large
enough, the overlap of wavefunctions on two different sites leads to a very strong exchange interaction, which 
is important, and could dominate the physics at Mn concentrations as low as 5\%.  There are two contributions
to this exchange interaction - superexchange, in which localized electrons are exchanged trough only one type of 
bands (either conduction or valence bands),   and the Bloumbergen-Rowland term, when the exchange is through both
conduction and valence bands.  We have found that, in case of direct band gap, the
Bloemberger-Rowland exchange is ferromagnetic. This could lead to a dramatic enhancement of Curie temperatures in
certain magnetic semiconductors with a direct band gap, such as GaMnAs.
 One other important consequence of the
indirect exchange is that, if it is ferromagnetic and reasonably long range, doped carriers are not necessary
to mediate ferromagnetism. This leads potentially to a new class of high-temperature magnetic semiconductors,
with high Curie temperatures determined entirely by the interaction between localized impurities,  
\textit{not Zener mechanism}.  This emphasizes the effort to search
for new materials, where ferromagnetism is not carrier-driven (for example, driven by the Bloembergen-Rowland mechanism).
Another important consequence of the Anderson model is that, if there are more than
one type of carriers (for example, light and heavy holes in GaAs), the long-range RKKY interaction becomes rather 
complicated, since it involves a ``mixed'' contribution. We have also found that, at large doping,
the RKKY interaction for the Anderson model and the $p-d$ model is  \textit{different} at short distances.
The effective exchange interaction in the $U=\infty$ Anderson model for any FS is given by Eqs.(\ref{exchangeint})
(\ref{exchch}),(\ref{exch}). A numerical solution of this effective Hamiltonian for a given set of parameters (determined
from quantum chemistry) would give the answer for $T_c(n_i)$ in the most general case.

We have briefly considered  effects of disorder and interactions. We have shown that when De Gennes\cite{deGennes}
approximation $p_F l \gg 1$ is applicable, disorder does not modify Curie temperature. Localization effects, however,
do modify Curie temperature, although their effect could be reduced to removal carriers. 
Finally, in this paper we have not considered the effects of mixing of conduction and valence bands (such as kp). These effects 
should also be included in the full description.
We should mention that application of the Anderson model to GaAs:Mn was also considered in Ref. \cite{Kikoin},
although the limits of the Anderson model and their conclusions are different from ours.

I would like to thank L. P. Gor'kov, Z. Fisk, and G. Kotliar for many useful discussions,
and H. Weitering for sharing his experimental observations. 
I am also very greatful to Misha Zhitomirsky for pointing out Ref.\cite{Kikoin}.
 
This work was supported by the University of Tennessee. I also acknowledge, with  gratitute, the 
input on this work provided by the participants of the Aspen 2003 winter conference on "Complex Quantum Order",
and Aspen Summer 2003 workshop on "Competing Orders and Quantum Criticality in Correlated Electrons, Bosons, and Spin Systems".

\end{document}